\documentclass[12pt,onecolumn,amsmath,amssymb]{revtex4-1}
\usepackage{graphicx}
\usepackage{dcolumn}
\usepackage{bm}

\begin{document}

\newcommand{\ymno}{YMnO$_{3}$}
\newcommand{\dyymno}{Dy$_{1-x}$Y$_{x}$MnO$_{3}$}
\newcommand{\dyy}{Dy$_{0.5}$Y$_{0.5}$MnO$_{3}$}
\newcommand{\erymno}{Er$_{1-x}$Y$_{x}$MnO$_{3}$}
\newcommand{\hoymno}{Ho$_{1-x}$Y$_{x}$MnO$_{3}$}
\newcommand{\hoermno}{Ho$_{1-x}$Er$_{x}$MnO$_{3}$}
\newcommand{\ermno}{ErMnO$_{3}$}
\newcommand{\homno}{HoMnO$_{3}$}
\newcommand{\lumno}{LuMnO$_{3}$}
\newcommand{\rmno}{\textit{R}MnO$_{3}$}
\newcommand{\neel}{N\'eel}
\newcommand{\mn}{Mn$^{3+}$}
\newcommand{\ho}{Ho$^{3+}$}
\newcommand{\dy}{Dy$^{3+}$}
\newcommand{\pone}{P6$^{'}_3$cm$^{'}$}        
\newcommand{\ptwo}{P6$^{'}_3$c$^{'}$m}        
\newcommand{\tsr}{T$_{SR}$}
\newcommand{\tn}{T$_{N}$}


\title{In-plane and out-of-plane correlations in \hoymno} 


\author{J. Gunasekera}
\email[]{tcszd5@mail.missouri.edu}
\author{O. P. Vajk}
\author{Y. Wang}
\author{K. Tarwater}
\affiliation{Department of Physics and Astronomy, University of Missouri, Columbia, Missouri 65211, USA}

\author{T. W Heitmann}
\affiliation{The Missouri Research Reactor, University of Missouri, Columbia, Missouri 65211, USA}


\date{\today}

\begin{abstract}
\newpage

Strong magnetic-ferroelectric coupling in hexagonal \homno~has been observed previously at spin reorientation temperature between \pone~and the \ptwo~magnetic phases.  In contrast, \ymno~has only a single magnetic phase (\pone) and no sign of strong magnetic-ferroelectric coupling.  In order to investigate the \pone~to \ptwo~spin reorientation transition, single crystals of \hoymno~at varying compositions were grown.  Neutron scattering measurements reveal quasielastic scattering in the \ymno-like phase (\ptwo) centered on the Bragg peak corresponding to \homno-like phase (\pone) and vice versa.  This scattering may be due to short-lived fluctuations into the "wrong" magnetic phase at domain boundaries.  These results suggest that there is strong phase competition between the \pone~and \ptwo~phases even in pure \ymno, and this competition may play an important role in the magnetic-ferroelectric coupling observed in \homno.


\end{abstract}

\pacs{}

\maketitle 


\newpage

\section{Introduction}

The hexagonal manganite multiferroics (\rmno, R = Rare earth) are a family of isostructural compounds which order ferroelectrically at high temperatures and antiferromagnetically at low temperatures \cite{fiebig03}.  Magnetic order in this family is created by 2 layers of \mn~stacked on top of each other at Z=0 and Z=1/2 with in-plane 120$^\circ$ angle between neighboring \mn~$S=2$ magnetic moments within each plane \cite{koehler64}.  Much of the interest in this family of multiferroics is due to the strong magnetic-ferroelectric coupling observed in \homno~\cite{lorenz04,yen05,hur09,lottermoser04}. As \homno~undergoes a spin reorientation transition between a \ptwo~magnetic phase and a \pone~phase with decreasing temperature \cite{lonkai02}, there is a very large increase in the \textit{c}-axis dielectric constant at the spin reorientation transition temperature \tsr~\cite{lorenz04,yen05} along with a change in the magnitude of the ferroelectric polarization \cite{hur09}. \homno~is the only pure member of this family of multiferroics where this strong magnetic-ferroelectric coupling at a spin reorientation transition has been observed, and the physics behind this strong coupling is not well understood.  In contrast, \ymno~orders antiferromagnetically in \pone~phase symmetry and does not undergo any spin reorientation at zero field \cite{fiebig03,lonkai02}.

The different magnetic phases in \rmno~are distinguished by how spins in neighboring planes are correlated, and the determining factor seems to be the position of the Mn$^{3+}$ ion within the unit cell respect to a critical value of 1/3 \cite{fabreges09}.  The mechanism responsible for the large magnetic-ferroelectric coupling observed in \homno~is still not well understood, though.  One approach to try to better understand this coupling is to study how various parameters change the spin reorientation temperature where this coupling is seen.  Such work has included magnetic fields \cite{fiebig03,lorenz04,yen05}, electric field \cite{lottermoser04}, pressure \cite{delacruz05}, and chemical substitution \cite{zhou05,zhou07,zhou08}.  By doping either Er or Y into \homno, the spin reorientation transition temperature \tsr~ can be decreased or increased, respectively.  This makes sense, since \ermno~orders in the \ptwo~phase and \ymno~orders in the \pone~phase.  However, we have previously found that the phase boundary between these two phases as a function of rare earth concentration is actually complex, with qualitatively different behavior for other R$_{1-x}$Y$_{x}$MnO$_{3}$ compounds \cite{vajk11,vajk12}.

Previous work done by H.D Zhou \textit{et al.} investigated the phase diagram of Ho$_{1-x}$Y$_{x}$MnO$_{3}$ with magnetic susceptibility, specific heat and dielectric constant  \cite{zhou07}.  For $x \geq 0.9$, Ho$_{1-x}$Y$_{x}$MnO$_{3}$ orders in the \pone~phase and no spin reorientation occurs.  They attribute the magnetic susceptibility and specific heat increases at \tsr~to partial ordering of \ho~moments, both of which are suppressed upon Y doping.  The dielectric susceptibility anomaly at \tsr~was observed for all samples with a spin reorientation transition, even up to $x=0.8$.  Pure \ymno~does not have any dielectric susceptibility anomalies along the \textit{c} axis \citep{katsufuji01}, not even the slight kink observed at the \neel~temperature \tn~in \homno~\citep{lorenz04}.  In order to further study the evolution of magnetic order in \hoymno, we have performed neutron scattering measurements of the spin dynamics in single-crystal samples.  Our results may help illuminate the mechanisms responsible for the coupling of magnetic and ferroelectric order in \homno.

\section{Experiment}

Single crystal samples of  \hoymno~were grown using an optical floating zone technique, where $x = 0, 0.65, 0.75, 0.825, 0.88, 1$. Powders of Ho$_2$O$_3$, Y$_2$O$_3$ and MnO$_2$ were mixed and ground together using a pestle and a mortar and calcinated at 1200 $^\circ$C for 8 hours. We repeated this process thrice to make it uniform. Afterwards we reground and packed in to both seed and feed rods and sintered at 1450 $^\circ$C for 8 hours. Crystals were grown in a 4-mirror floating zone furnace under air at atmospheric pressure. Grown crystals were approximately 2 to 5 grams. Neutrons scattering of single crystal samples were performed at the Missouri University Research Reactor (MURR) using the triple-axis spectrometer TRIAX, using a closed-cycle helium displex to cool the samples to low temperature.

\subsection{Elastic Neutron Scattering}

\begin{figure}[tbp]
\includegraphics[width=8cm]{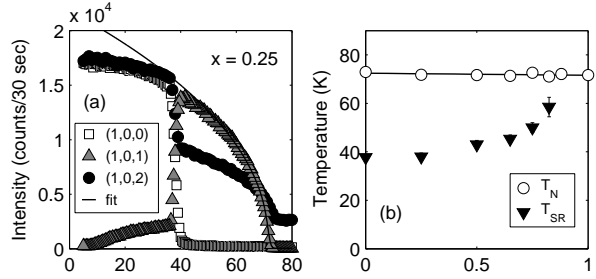}%
\caption{\label{OP}
Order parameters for Ho$_{0.75}$Y$_{0.25}$MnO$_{3}$ concentration.  In the left panel, white squares show the (1,0,0) peak intensity, gray triangles show the (1,0,1) peak intensity, and black circles shows the intensity of (1,0,2) peak.  The (1,0,0) reflection is associated with the \ptwo~phase, the (1,0,1) reflection is associated with the \pone~phase, and the (1,0,2) relfection is associated with both phases but at different intensities.  Lines show fits to the (1,0,1) order parameter data near \tn.
The right panel show the phase diagram extracted from order parameter measurements.  White circles indicate \tn, black triangles indicate \tsr.}
\end{figure}

When the spins undergo 90$^\circ$ reorientation from \ptwo~to \pone~there is a change in intensity of the magnetic Bragg reflections.  In the \ptwo~phase between \tn~and \tsr~the (1,0,1) Bragg reflection dominates, while in the \pone~phase below \tsr~and in samples with no spin reorientation the (1,0,0) Bragg reflection dominates \cite{munoz01}.  The (1,0,2) reflection is both structural and magnetic, and has significant scattering but different intensities in both magnetic phases.  Figure \ref{OP}~shows example order parameter scans for Ho$_{0.75}$Y$_{0.25}$MnO$_{3}$.  The Bragg scattering intensity scales with the square of the antiferromagnetic order parameter M$_{ST}$. The \neel~temperature \tn~was obtained by fitting the peak intensity  \textit{I} of the dominant reflection above 60 K to the form of a second-order phase transition:
\begin{eqnarray}  
I \propto M_{st}^2 \propto (T_N - T)^{2\beta}.
\end{eqnarray}
The fits include a small Gaussian spread in the transition temperature (shown by the error bars for \tn~).  The critical exponent $\beta$ value varied from 0.187 to 0.333 and was very sensitive to the fitting range of data.  These critical exponent values should not be considered as the true critical exponents of the \neel~transition, but \tn~obtained from fitting can be considered reliable since they were not very sensitive to the fitting range.  The shift in scattering intensity from the (1,0,1) to the (1,0,0) with decreasing temperature indicates \tsr, and error bars indicate the width of this first-order transition.  With increasing Yttrium concentration, \tsr~increases towards to \tn~ and eventually at $x=0.88$ \tsr~completely vanishes.  Results for \tn~and \tsr~are shown in Figure \ref{OP}.  The magnetic phase diagram for \hoymno~we obtained is consistent with previous work done by Zhou \textit{et al.} \cite{zhou07}.

\subsection{Inelastic Neutron Scattering}

Below the \neel~temperature, inelastic measurements of \ymno~revealed quasielastic scattering mainly around (1,0,1), and secondary scattering (1,0,2) Bragg positions.  Example data taken at 30K are shown in Figure \ref{lorVSGauss}, which revealed a peak centered at zero energy transfer, but with an energy width much larger than instrumental resolution.  Scans along the $H$ and $L$ directions (in-plane and out-of-plane, respectively) were performed at a constant energy transfer of 0.7 meV in order to reduce elastic Bragg scattering.  Previous work done by Sato \textit{et al.} \cite{sato03} reported quasielastic scattering with similar in-plane $H$ momentum transfers, but with no observed out-of-plane momentum dependence.

\begin{figure}[tbp]
\includegraphics[width=8cm]{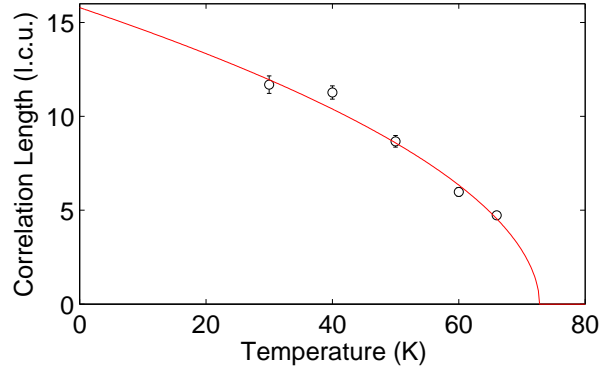}%
\caption{\label{L_correlations}
Out-of-plane correlation length for quasielastic scattering in \ymno~versus temperature.  The line is a fit to the heuristic form described in the text.}
\end{figure}

The data were fit to a structure factor $S({\bf Q},\omega)$ convoluted with the spectrometer resolution, where the structure factor model was given by the following equation:
\begin{eqnarray}  
S({\bf Q},\omega) \propto \frac{1}{1-e^{-\hbar \omega/k_{B}T}} \frac{\Gamma \omega}{\Gamma^2 + \omega^2} \, \frac{W_{\parallel}^2}{\Delta H^2+W_{\parallel}^2 }  \frac{W_{\parallel}^2}{\Delta K^2+W_{\parallel}^2} e^ \frac{-\Delta L^2}{(W_{\perp})^2}.
\end{eqnarray}
$W_{\parallel}$ is the in-plane width of the scattering, $W_{\perp}$ is the out-of-plane width, and $\Gamma$ is the energy width. $\Delta H$, $\Delta K$, and $\Delta L$ are the difference in H, K, and L respectively from the center of the quasielastic scattering.  The first fraction is the Boltzmann term, followed by Lorentzians in energy and in-plane momenta $H$ and $K$.  The last term is a Gaussian in out-of-plane momentum $L$.  Figure \ref{lorVSGauss}(c) shows fits using both the above static structure factor (solid line) and a similar form using a Lorentzian instead of a Gaussian for the out-of-plane dependence, and the Gaussian fits produced visibly better results.  The out-of-plane correlation length in lattice units is given by $2/W_{\perp}$, and results are plotted in Figure \ref{L_correlations}.  The data were fit to the heuristic form 
\begin{eqnarray}  
I  \propto (T_N - T)^{2\lambda}.
\end{eqnarray}
where $\lambda \approx 0.26$ for the above fit.  With increasing temperature the out-of-plane correlation length decreases, and above \tn~no apparent out-of-plane correlations persist.  The total quasielastic scattering intensity also decreases with decreasing temperature below \tn.

\begin{figure}[tbp]
\includegraphics[width=8cm]{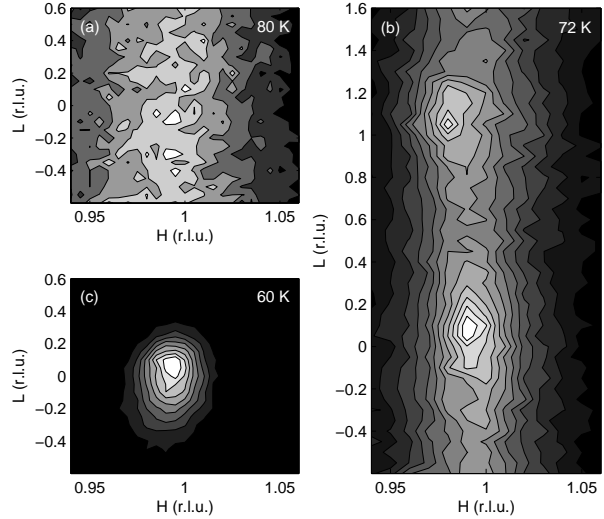}%
\caption{\label{HvT}
Q$_{L}$ v Q$_{H}$ scattering intensity maps for Ho$_{0.25}$Y$_{0.75}$MnO$_{3}$  at a constant 0.5 meV energy transfer. Scans were taken (a) at 80K (above \tn), (b) 72K (approximately \tn) and (c) 60K (below \tn).  Above \tn, the quasielastic scattering shows no out-of-plane dependence.  At \tn, the quasielastic scattering begins to concentrate around both the (1,0,0) and (1,0,1) positions.  At 60K, the scattering is centered around the (1,0,0) position, although the magnetic Bragg scattering is located at the (1,0,1) position at this temperature.
}
\end{figure}

Similar quasielastic scattering was also observed in \hoymno~samples for $0 < x < 1$.  Figure \ref{HvT} shows scattering intensity maps as a function of \textit{H} and \textit{L} at three different temperatures for Ho$_{0.25}$Y$_{0.75}$MnO$_{3}$.  Measurements were taken at a constant energy transfer of 0.5 meV to reduce elastic background scattering.  Above \tn, the scattering was broad in the $H$ direction and had no $L$ dependence, as shown in Figure \ref{HvT}(a), indicating short-range in-plane correlations with no out-of-plane correlations.  Near \tn, the quasielastic scattering showed two quasielstic peaks (Figure \ref{HvT}(b)) centered at the (1,0,0) and (1,0,1) positions. Below \tn, scattering became sharper in both in plane and out of plane directions, and was only centered around the (1,0,0) position at 60K (Figure \ref{HvT}(c)).  

\begin{figure}[tbp]
\includegraphics[width=8cm]{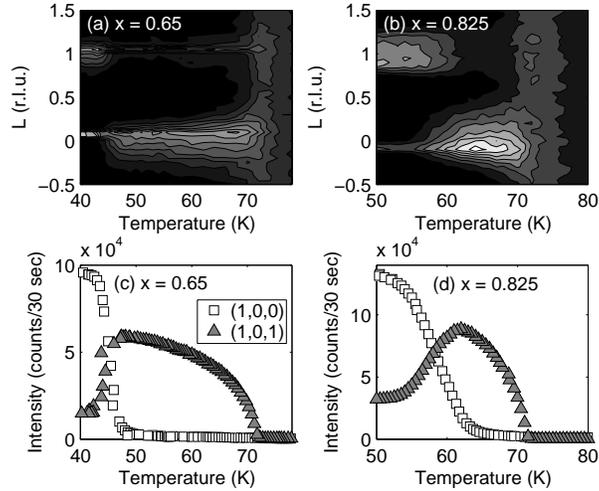}%
\caption{\label{LvT}
Top panels: Measurements of the out-of-plane dependence of quasielastic scattering versus temperature for two compositions of \hoymno.  Bottom panels: Order parameter measurements of the same samples over the same temperature range.  At the spin reorientation temperature, elastic Bragg scattering shifts from the (1,0,1) position to the (1,0,0) position upon cooling, while quasielastic scattering shifts from the (1,0,0) position to the (1,0,1) position at the same temperature.  Scattering with very narrow L width is also visible in the top panel, and is due to the tail end of the resolution ellipsoid clipping the Bragg reflection}
\end{figure}

For \hoymno~samples, which undergo a spin reorientation transition, the location of the quasielastic peak was observed to shift at this transition.  Figure \ref{LvT} illustrates this shift from (1,0,0) to (1,0,1) at \tsr~for two different concentrations of \hoymno.  When the system is in the \pone~phase below \tsr, the magnetic Bragg reflection is located at (1,0,0) (lower panel) and the quasielastic scattering is located at (1,0,1)(upper panel). Above \tsr~the magnetic Bragg reflection is located at (1,0,1) and the quasielastic scattering shifts to the (1,0,0) position.  Quasielastic scattering was also observable at the (1,0,2) position in both phases.  These results match similar observations in \erymno~and \dyymno~\cite{vajk11,vajk12}.

\section{Discussion}

Diffuse neutron scattering in \rmno~similar to our observations has been previously reported in several papers \cite{sato03,park03,Roessli05,sekhar05,Lewtas10}.  Sato \textit{et al.} reported quasielastic scattering above and below \tn~in pure \ymno, with an in-plane width that became resolution limited below \tn, an energy width that remained wider than resolution down to the lowest temperatures, but with no reported out-of-plane correlation \cite{sato03}.  Park \textit{et al.} also reported quasielastic scattering above and below \tn~in \ymno, but their samples were powder and so the data does not provide directional information\cite{park03}.  Roessli \textit{et al.} measured quasielastic scattering in pure \ymno~around the (1,0,0) position above \tn~but not below \cite{Roessli05}.  Far above \tn, the scattering had no out-of-plane correlations, but near \tn~the scattering developed $L$ dependence with a peak centered around (1,0,0), and a peak shape well-described by a Lorentzian in both the in-plane and out-of-plane direction.  Polarized neutron measurements indicated that the quasielastic scattering was due to in-plane fluctuations.  Lewtas \textit{et al.} reported diffuse elastic scattering in \lumno~around the (1,0,0) position above \tn, while the primary magnetic Bragg reflection below \tn~was at the (1,0,1) position \citep{Lewtas10}.  This scattering may have been quasielastic, but no energy-dependent measurements on the diffuse scattering was reported.  Sekhar \textit{et al.} reported diffuse scattering below \tn~in powder neutron measurements of \erymno, but only in samples with high Yttrium concentrations \cite{sekhar05}.  This scattering occurs at $2\theta$ angles similar to the (1,0,0) and (1,0,1) peaks, but they did not identify which of those two positions this scattering was associated with.  

\begin{figure}[t]
	\includegraphics[width=8cm]{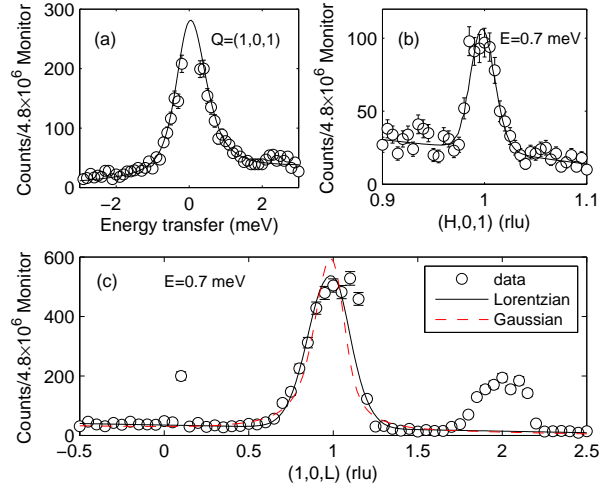}
	\caption[Fit for \ymno~at 30K]{\label{lorVSGauss} Measurements of the quasielastic scattering in \ymno~at 40K.  Scattering is centered at the Q=(1,0,1) position, and scans were taken with varying energy (a), in-plane momentum (b), and out-of-plane momentum (c).  Momentum scans were taken with fixed energy transfer of 0.7 meV to reduce elastic scattering background.  Note that quasielastic scattering is also present around the (1,0,2) position.  Lines are fits to the data described in the text. The solid line in (c) shows the fit result using a Gaussian for the out-of-plane dependence, while the dashed line shows the fit result using a Lorentzian.}
\end{figure}

There is agreement among all the single-crystal measurements that well above \tn, the diffuse/quasielastic scattering has no out-of-plane correlations and very short in-plane correlations.  Both Lewtas \textit{et al.} and Roessli \textit{et al.} report that the scattering develops $L$ dependence centered around the (1,0,0) position in \lumno~and \ymno, respectively, but neither report measuring around the (1,0,1)\cite{Lewtas10,Roessli05}.  This omission is significant because the (1,0,1) position is the magnetic Bragg position for \lumno, but the (1,0,0) is the magnetic Bragg position for \ymno.  Below the \neel~temperature, there is significant discrepency between our work and the work of Sato \textit{et al.}\cite{sato03}, since the latter report that the quasielastic scattering has no out-of-plane dependence, while we observe very strong out-of-plane dependence below \tn.  We speculate that the Sato \textit{et al.} measurements may simply not have extended far enough.  Only one $L$ scan was shown, and the range covered, from $L=-0.2$ to $L=0.5$, corresponds to an essentially flat region of our scans, as seen in Figure \ref{lorVSGauss}.  Our own work on \hoymno, as well as our previous work on \erymno~and \dyymno, has shown that there are significant out-of-plane correlations in the quasielastic scattering of \rmno~below \tn. However, while these quasielastic excitations have now been observed in \ymno~\cite{sato03,park03,Roessli05} and \lumno~\cite{Lewtas10}, as well as \hoymno, \erymno~ \cite{sekhar05,vajk11}, and \dyymno~\cite{vajk12} at high Yttrium concentrations, they were not observed in inelastic neutron scattering measurements of pure \homno~\cite{vajk06}~or in pure \ermno~ diffraction measurments \cite{sekhar05}.

Well above \tn, the quasielastic scattering in these materials shows no sign of \textit{L} dependence.  Correlations are in-plane only, and so no distinction can be made between phases which differ in their out-of-plane stacking arrangement.  Near \tn, however, \textit{L} dependence begins to emerge as correlations begin to develop between MnO planes.  The presence of quasielastic scattering at both the (1,0,0) and (1,0,1) positions near the \tn~indicates that correlated regions of both phases exist above \tn.  Neither phase is strongly preferred, but instead they appear to compete with each other. According to our interpretation there should be a peak in the quasielastic scattering just above \tn~at the (1,0,1) position in \lumno~in addition to the observed peak around (1,0,0) \citep{Lewtas10}, but it may simply not have been measured.
 
Below \tn, \ymno~orders in only one phase, and quasielastic scattering for the ordered phase disappears as the correlations become static.  But quasielastic scattering for the ``wrong" phase persist below \tn.  These patches are broad in-plane and narrow out-of-plane.  We hypothesize that the quasielastic scattering originates from domain boundaries within our crystal.  Mn$^{3+}$ moments located between different domains will be frustrated.  At \textit{c}-axis domain boundaries, the weak, frustrated out-of-plane interactions may prevent long-range static order, but in-plane correlations within those boundaries could still lead to short-range correlated fluctuations.  Thermal fluctuations may excite spins at the boundary into the wrong magnetic phase, which still satisfies the in-plane interactions.  If the correlations which form are short-lived, they will lead to quasielastic scattering.  Because the in-plane coupling is so much stronger than the out-of-plane coupling, fluctuations at in-plane domain boundaries may not be able to form correlated patches of any significant size, and so would not contribute to the observed quasielastic scattering.  This would also explain the line shape of the quasielastic scattering: the shape of the domain walls would constrain correlations along the \textit{c} axis, but correlations could grow to much larger size along the in-plane direction.  The neighboring ordered domains will not only act to limit the size of the out-of-plane correlation length, they may also cause those correlations to decay faster than an exponential.  A Lorentzian line shape indicates correlations which decay exponentially with distance, while a Gaussian indicates correlations which decay much faster than exponential.  This contrasts with the analysis of Roessli \textit{et al.}, who used a Lorentzian to analyze the $L$ dependence of the quasielastic scattering above \tn~\cite{Roessli05}.  Without an ordered phase to constrain the fluctuations, correlations could extend as a decaying exponential along the $c$ axis, so their model is still consistent with our hypothesis.

We also hypothesize that the existence of quasielastic scattering in Yttrium-rich \rmno~but not pure \homno~is due to the larger in-plane easy-axis anisotropy in \homno.  This in-plane anisotropy for \mn~moments gives rise to a spin gap which was measured by Fabreges \textit{et al.} for both Yttrium and Holmium, who found a significantly larger gap for \homno~than for \ymno~\cite{fabreges09}.  This anisotropy acts as an energy barrier to in-plane rotation, and so a large anisotropy may prevent the quasielastic fluctuations at the domain boundaries.  Further evidence for this interpretation comes from the polarized scattering measurements of Roessli \textit{et al.}, who found that the quasielastic scattering is due to spins fluctuating within the plane \citep{Roessli05}.  Out-of-plane fluctuations gave rise to inelastic scattering due to the easy-plane anisotropy, which is much larger than the in-plane anisotropy for all \rmno.  The observation of quasielastic scattering in \lumno~is consistent with this interpretation, since the spin-wave dispersion measured by Lewtas \textit{et al.} had a vanishingly small gap indicating very little in-plane anisotropy \cite{Lewtas10}.

One possible objection to our interpretation is that the observed quasielastic scattering may simply be the spin-wave dispersion approaching zero energy.  The spin-wave measurements by Fabreges \textit{et al.} \cite{fabreges09} do indicate that the dispersion is close to gapless for \ymno~at the (1,0,1) position, and in \homno~it has a minimum at either the (1,0,1) or the (1,0,0), depending on the temperature, in the same place as the observed quasielastic scattering for \ymno~and \hoymno.  However, the observation of quasielastic scattering at the (1,0,2), where the dispersion should move to higher energy in \ymno~due to out-of-plane coupling, indicates that the quasielastic scattering is not simply a measurement artifact of a gapless dispersion, but is instead related to correlations matching an ordered magnetic phase, albeit the wrong phase.

Our model of quasielastic scattering in \rmno~explains both our own observations as well as a number of disparate previous results.  It also explains the \textit{absence} of quasielastic scattering for \homno.  If our model is correct, then other \rmno~compounds with large in-plane anisotropy, such as YbMnO$_{3}$ \cite{fabreges09}, should not exhibit quasielastic scattering either.  This scattering indicates strong competition between competing 3D-ordered phases.  It is particularly interesting that even though \homno~exhibits both \pone~and \ptwo~phases, quasielastic scattering is absent.  The in-plane anisotropy appears to stabilize \textit{both} magnetic phases in \homno.  This suggests that the spin reorientation transition in \homno~is driven by something besides purely magnetic interactions.  We know that the magnetism and the ferroelectricity are coupled at this transition, and a better understanding of the dynamics involved may help illuminate the nature of this magnetic-ferroelectric interaction.

\begin{acknowledgments}
The work was partly supported by NSF IGERT Grant No. DGE-1069091.
\end{acknowledgments}


\newpage


\begin{thebibliography}{22}%
\makeatletter
\providecommand \@ifxundefined [1]{%
 \@ifx{#1\undefined}
}%
\providecommand \@ifnum [1]{%
 \ifnum #1\expandafter \@firstoftwo
 \else \expandafter \@secondoftwo
 \fi
}%
\providecommand \@ifx [1]{%
 \ifx #1\expandafter \@firstoftwo
 \else \expandafter \@secondoftwo
 \fi
}%
\providecommand \natexlab [1]{#1}%
\providecommand \enquote  [1]{``#1''}%
\providecommand \bibnamefont  [1]{#1}%
\providecommand \bibfnamefont [1]{#1}%
\providecommand \citenamefont [1]{#1}%
\providecommand \href@noop [0]{\@secondoftwo}%
\providecommand \href [0]{\begingroup \@sanitize@url \@href}%
\providecommand \@href[1]{\@@startlink{#1}\@@href}%
\providecommand \@@href[1]{\endgroup#1\@@endlink}%
\providecommand \@sanitize@url [0]{\catcode `\\12\catcode `\$12\catcode
  `\&12\catcode `\#12\catcode `\^12\catcode `\_12\catcode `\%12\relax}%
\providecommand \@@startlink[1]{}%
\providecommand \@@endlink[0]{}%
\providecommand \url  [0]{\begingroup\@sanitize@url \@url }%
\providecommand \@url [1]{\endgroup\@href {#1}{\urlprefix }}%
\providecommand \urlprefix  [0]{URL }%
\providecommand \Eprint [0]{\href }%
\providecommand \doibase [0]{http://dx.doi.org/}%
\providecommand \selectlanguage [0]{\@gobble}%
\providecommand \bibinfo  [0]{\@secondoftwo}%
\providecommand \bibfield  [0]{\@secondoftwo}%
\providecommand \translation [1]{[#1]}%
\providecommand \BibitemOpen [0]{}%
\providecommand \bibitemStop [0]{}%
\providecommand \bibitemNoStop [0]{.\EOS\space}%
\providecommand \EOS [0]{\spacefactor3000\relax}%
\providecommand \BibitemShut  [1]{\csname bibitem#1\endcsname}%
\let\auto@bib@innerbib\@empty
\bibitem [{\citenamefont {Fiebig}\ \emph {et~al.}(2003)\citenamefont {Fiebig},
  \citenamefont {Lottermoser},\ and\ \citenamefont {Pisarev}}]{fiebig03}%
  \BibitemOpen
  \bibfield  {author} {\bibinfo {author} {\bibfnamefont {M.}~\bibnamefont
  {Fiebig}}, \bibinfo {author} {\bibfnamefont {T.}~\bibnamefont {Lottermoser}},
  \ and\ \bibinfo {author} {\bibfnamefont {R.~V.}\ \bibnamefont {Pisarev}},\
  }\href {\doibase 10.1063/1.1544513} {\bibfield  {journal} {\bibinfo
  {journal} {Journal of Applied Physics}\ }\textbf {\bibinfo {volume} {93}},\
  \bibinfo {pages} {8194} (\bibinfo {year} {2003})}\BibitemShut {NoStop}%
\bibitem [{\citenamefont {Koehler}\ \emph {et~al.}(1964)\citenamefont
  {Koehler}, \citenamefont {Yakel}, \citenamefont {Wollan},\ and\ \citenamefont
  {Cable}}]{koehler64}%
  \BibitemOpen
  \bibfield  {author} {\bibinfo {author} {\bibfnamefont {W.~C.}\ \bibnamefont
  {Koehler}}, \bibinfo {author} {\bibfnamefont {H.~L.}\ \bibnamefont {Yakel}},
  \bibinfo {author} {\bibfnamefont {E.~O.}\ \bibnamefont {Wollan}}, \ and\
  \bibinfo {author} {\bibfnamefont {J.~W.}\ \bibnamefont {Cable}},\ }\href@noop
  {} {\bibfield  {journal} {\bibinfo  {journal} {Phys. Lett.}\ }\textbf
  {\bibinfo {volume} {9}},\ \bibinfo {pages} {93} (\bibinfo {year}
  {1964})}\BibitemShut {NoStop}%
\bibitem [{\citenamefont {Lorenz}\ \emph {et~al.}(2004)\citenamefont {Lorenz},
  \citenamefont {Litvinchuk}, \citenamefont {Gospodinov},\ and\ \citenamefont
  {Chu}}]{lorenz04}%
  \BibitemOpen
  \bibfield  {author} {\bibinfo {author} {\bibfnamefont {B.}~\bibnamefont
  {Lorenz}}, \bibinfo {author} {\bibfnamefont {A.~P.}\ \bibnamefont
  {Litvinchuk}}, \bibinfo {author} {\bibfnamefont {M.~M.}\ \bibnamefont
  {Gospodinov}}, \ and\ \bibinfo {author} {\bibfnamefont {C.~W.}\ \bibnamefont
  {Chu}},\ }\href {\doibase 10.1103/PhysRevLett.92.087204} {\bibfield
  {journal} {\bibinfo  {journal} {Phys. Rev. Lett.}\ }\textbf {\bibinfo
  {volume} {92}},\ \bibinfo {pages} {087204} (\bibinfo {year}
  {2004})}\BibitemShut {NoStop}%
\bibitem [{\citenamefont {Yen}\ \emph {et~al.}(2005)\citenamefont {Yen},
  \citenamefont {dela Cruz}, \citenamefont {Lorenz}, \citenamefont {Sun},
  \citenamefont {Wang}, \citenamefont {Gospodinov},\ and\ \citenamefont
  {Chu}}]{yen05}%
  \BibitemOpen
  \bibfield  {author} {\bibinfo {author} {\bibfnamefont {F.}~\bibnamefont
  {Yen}}, \bibinfo {author} {\bibfnamefont {C.~R.}\ \bibnamefont {dela Cruz}},
  \bibinfo {author} {\bibfnamefont {B.}~\bibnamefont {Lorenz}}, \bibinfo
  {author} {\bibfnamefont {Y.~Y.}\ \bibnamefont {Sun}}, \bibinfo {author}
  {\bibfnamefont {Y.~Q.}\ \bibnamefont {Wang}}, \bibinfo {author}
  {\bibfnamefont {M.~M.}\ \bibnamefont {Gospodinov}}, \ and\ \bibinfo {author}
  {\bibfnamefont {C.~W.}\ \bibnamefont {Chu}},\ }\href {\doibase
  10.1103/PhysRevB.71.180407} {\bibfield  {journal} {\bibinfo  {journal} {Phys.
  Rev. B}\ }\textbf {\bibinfo {volume} {71}},\ \bibinfo {pages} {180407}
  (\bibinfo {year} {2005})}\BibitemShut {NoStop}%
\bibitem [{\citenamefont {Hur}\ \emph {et~al.}(2009)\citenamefont {Hur},
  \citenamefont {Jeong}, \citenamefont {Hundley}, \citenamefont {Kim},\ and\
  \citenamefont {Cheong}}]{hur09}%
  \BibitemOpen
  \bibfield  {author} {\bibinfo {author} {\bibfnamefont {N.}~\bibnamefont
  {Hur}}, \bibinfo {author} {\bibfnamefont {I.~K.}\ \bibnamefont {Jeong}},
  \bibinfo {author} {\bibfnamefont {M.~F.}\ \bibnamefont {Hundley}}, \bibinfo
  {author} {\bibfnamefont {S.~B.}\ \bibnamefont {Kim}}, \ and\ \bibinfo
  {author} {\bibfnamefont {S.-W.}\ \bibnamefont {Cheong}},\ }\href {\doibase
  10.1103/PhysRevB.79.134120} {\bibfield  {journal} {\bibinfo  {journal} {Phys.
  Rev. B}\ }\textbf {\bibinfo {volume} {79}},\ \bibinfo {pages} {134120}
  (\bibinfo {year} {2009})}\BibitemShut {NoStop}%
\bibitem [{\citenamefont {Lottermoser}\ \emph {et~al.}(2004)\citenamefont
  {Lottermoser}, \citenamefont {Lonkai}, \citenamefont {Amann}, \citenamefont
  {Hohlwein}, \citenamefont {Ihringer},\ and\ \citenamefont
  {Fiebig}}]{lottermoser04}%
  \BibitemOpen
  \bibfield  {author} {\bibinfo {author} {\bibfnamefont {T.}~\bibnamefont
  {Lottermoser}}, \bibinfo {author} {\bibfnamefont {T.}~\bibnamefont {Lonkai}},
  \bibinfo {author} {\bibfnamefont {U.}~\bibnamefont {Amann}}, \bibinfo
  {author} {\bibfnamefont {D.}~\bibnamefont {Hohlwein}}, \bibinfo {author}
  {\bibfnamefont {J.}~\bibnamefont {Ihringer}}, \ and\ \bibinfo {author}
  {\bibfnamefont {M.}~\bibnamefont {Fiebig}},\ }\href@noop {} {\bibfield
  {journal} {\bibinfo  {journal} {Nature}\ }\textbf {\bibinfo {volume} {430}},\
  \bibinfo {pages} {541} (\bibinfo {year} {2004})}\BibitemShut {NoStop}%
\bibitem [{\citenamefont {Lonkai}\ \emph {et~al.}(2002)\citenamefont {Lonkai},
  \citenamefont {Hohlwein}, \citenamefont {Ihringer},\ and\ \citenamefont
  {Prandl}}]{lonkai02}%
  \BibitemOpen
  \bibfield  {author} {\bibinfo {author} {\bibfnamefont {T.}~\bibnamefont
  {Lonkai}}, \bibinfo {author} {\bibfnamefont {D.}~\bibnamefont {Hohlwein}},
  \bibinfo {author} {\bibfnamefont {J.}~\bibnamefont {Ihringer}}, \ and\
  \bibinfo {author} {\bibfnamefont {W.}~\bibnamefont {Prandl}},\ }\href@noop {}
  {\bibfield  {journal} {\bibinfo  {journal} {Appl. Phys. A}\ }\textbf
  {\bibinfo {volume} {74}},\ \bibinfo {pages} {S843} (\bibinfo {year}
  {2002})}\BibitemShut {NoStop}%
\bibitem [{\citenamefont {Fabr\`eges}\ \emph {et~al.}(2009)\citenamefont
  {Fabr\`eges}, \citenamefont {Petit}, \citenamefont {Mirebeau}, \citenamefont
  {Pailh\`es}, \citenamefont {Pinsard}, \citenamefont {Forget}, \citenamefont
  {Fernandez-Diaz},\ and\ \citenamefont {Porcher}}]{fabreges09}%
  \BibitemOpen
  \bibfield  {author} {\bibinfo {author} {\bibfnamefont {X.}~\bibnamefont
  {Fabr\`eges}}, \bibinfo {author} {\bibfnamefont {S.}~\bibnamefont {Petit}},
  \bibinfo {author} {\bibfnamefont {I.}~\bibnamefont {Mirebeau}}, \bibinfo
  {author} {\bibfnamefont {S.}~\bibnamefont {Pailh\`es}}, \bibinfo {author}
  {\bibfnamefont {L.}~\bibnamefont {Pinsard}}, \bibinfo {author} {\bibfnamefont
  {A.}~\bibnamefont {Forget}}, \bibinfo {author} {\bibfnamefont {M.~T.}\
  \bibnamefont {Fernandez-Diaz}}, \ and\ \bibinfo {author} {\bibfnamefont
  {F.}~\bibnamefont {Porcher}},\ }\href {\doibase
  10.1103/PhysRevLett.103.067204} {\bibfield  {journal} {\bibinfo  {journal}
  {Phys. Rev. Lett.}\ }\textbf {\bibinfo {volume} {103}},\ \bibinfo {pages}
  {067204} (\bibinfo {year} {2009})}\BibitemShut {NoStop}%
\bibitem [{\citenamefont {dela Cruz}\ \emph {et~al.}(2005)\citenamefont {dela
  Cruz}, \citenamefont {Yen}, \citenamefont {Lorenz}, \citenamefont {Wang},
  \citenamefont {Sun}, \citenamefont {Gospodinov},\ and\ \citenamefont
  {Chu}}]{delacruz05}%
  \BibitemOpen
  \bibfield  {author} {\bibinfo {author} {\bibfnamefont {C.}~\bibnamefont {dela
  Cruz}}, \bibinfo {author} {\bibfnamefont {F.}~\bibnamefont {Yen}}, \bibinfo
  {author} {\bibfnamefont {B.}~\bibnamefont {Lorenz}}, \bibinfo {author}
  {\bibfnamefont {Y.~Q.}\ \bibnamefont {Wang}}, \bibinfo {author}
  {\bibfnamefont {Y.~Y.}\ \bibnamefont {Sun}}, \bibinfo {author} {\bibfnamefont
  {M.~M.}\ \bibnamefont {Gospodinov}}, \ and\ \bibinfo {author} {\bibfnamefont
  {C.~W.}\ \bibnamefont {Chu}},\ }\href {\doibase 10.1103/PhysRevB.71.060407}
  {\bibfield  {journal} {\bibinfo  {journal} {Phys. Rev. B}\ }\textbf {\bibinfo
  {volume} {71}},\ \bibinfo {pages} {060407} (\bibinfo {year}
  {2005})}\BibitemShut {NoStop}%
\bibitem [{\citenamefont {Zhou}\ \emph {et~al.}(2005)\citenamefont {Zhou},
  \citenamefont {Denyszyn},\ and\ \citenamefont {Goodenough}}]{zhou05}%
  \BibitemOpen
  \bibfield  {author} {\bibinfo {author} {\bibfnamefont {H.~D.}\ \bibnamefont
  {Zhou}}, \bibinfo {author} {\bibfnamefont {J.~C.}\ \bibnamefont {Denyszyn}},
  \ and\ \bibinfo {author} {\bibfnamefont {J.~B.}\ \bibnamefont {Goodenough}},\
  }\href {\doibase 10.1103/PhysRevB.72.224401} {\bibfield  {journal} {\bibinfo
  {journal} {Phys. Rev. B}\ }\textbf {\bibinfo {volume} {72}},\ \bibinfo
  {pages} {224401} (\bibinfo {year} {2005})}\BibitemShut {NoStop}%
\bibitem [{\citenamefont {Zhou}\ \emph {et~al.}(2007)\citenamefont {Zhou},
  \citenamefont {Lu}, \citenamefont {Vasic}, \citenamefont {Vogt},
  \citenamefont {Janik}, \citenamefont {Brooks},\ and\ \citenamefont
  {Wiebe}}]{zhou07}%
  \BibitemOpen
  \bibfield  {author} {\bibinfo {author} {\bibfnamefont {H.~D.}\ \bibnamefont
  {Zhou}}, \bibinfo {author} {\bibfnamefont {J.}~\bibnamefont {Lu}}, \bibinfo
  {author} {\bibfnamefont {R.}~\bibnamefont {Vasic}}, \bibinfo {author}
  {\bibfnamefont {B.~W.}\ \bibnamefont {Vogt}}, \bibinfo {author}
  {\bibfnamefont {J.~A.}\ \bibnamefont {Janik}}, \bibinfo {author}
  {\bibfnamefont {J.~S.}\ \bibnamefont {Brooks}}, \ and\ \bibinfo {author}
  {\bibfnamefont {C.~R.}\ \bibnamefont {Wiebe}},\ }\href {\doibase
  10.1103/PhysRevB.75.132406} {\bibfield  {journal} {\bibinfo  {journal} {Phys.
  Rev. B}\ }\textbf {\bibinfo {volume} {75}},\ \bibinfo {pages} {132406}
  (\bibinfo {year} {2007})}\BibitemShut {NoStop}%
\bibitem [{\citenamefont {Zhou}\ \emph {et~al.}(2008)\citenamefont {Zhou},
  \citenamefont {Vasic}, \citenamefont {Lu}, \citenamefont {Brooks},\ and\
  \citenamefont {Wiebe}}]{zhou08}%
  \BibitemOpen
  \bibfield  {author} {\bibinfo {author} {\bibfnamefont {H.~D.}\ \bibnamefont
  {Zhou}}, \bibinfo {author} {\bibfnamefont {R.}~\bibnamefont {Vasic}},
  \bibinfo {author} {\bibfnamefont {J.}~\bibnamefont {Lu}}, \bibinfo {author}
  {\bibfnamefont {J.~S.}\ \bibnamefont {Brooks}}, \ and\ \bibinfo {author}
  {\bibfnamefont {C.~R.}\ \bibnamefont {Wiebe}},\ }\href
  {http://stacks.iop.org/0953-8984/20/i=3/a=035211} {\bibfield  {journal}
  {\bibinfo  {journal} {Journal of Physics: Condensed Matter}\ }\textbf
  {\bibinfo {volume} {20}},\ \bibinfo {pages} {035211} (\bibinfo {year}
  {2008})}\BibitemShut {NoStop}%
\bibitem [{\citenamefont {Vajk}\ \emph {et~al.}(2011)\citenamefont {Vajk},
  \citenamefont {Gunasekera}, \citenamefont {Wang},\ and\ \citenamefont
  {Heitmann}}]{vajk11}%
  \BibitemOpen
  \bibfield  {author} {\bibinfo {author} {\bibfnamefont {O.~P.}\ \bibnamefont
  {Vajk}}, \bibinfo {author} {\bibfnamefont {J.}~\bibnamefont {Gunasekera}},
  \bibinfo {author} {\bibfnamefont {Y.}~\bibnamefont {Wang}}, \ and\ \bibinfo
  {author} {\bibfnamefont {T.}~\bibnamefont {Heitmann}},\ }\href {\doibase
  10.1063/1.3554263} {\bibfield  {journal} {\bibinfo  {journal} {Journal of
  Applied Physics}\ }\textbf {\bibinfo {volume} {109}},\ \bibinfo {eid}
  {07D910} (\bibinfo {year} {2011})}\BibitemShut {NoStop}%
\bibitem [{\citenamefont {Vajk}\ \emph {et~al.}(2012)\citenamefont {Vajk},
  \citenamefont {Wang}, \citenamefont {Gunasekera}, \citenamefont {Tarwater},\
  and\ \citenamefont {Heitmann}}]{vajk12}%
  \BibitemOpen
  \bibfield  {author} {\bibinfo {author} {\bibfnamefont {O.~P.}\ \bibnamefont
  {Vajk}}, \bibinfo {author} {\bibfnamefont {Y.}~\bibnamefont {Wang}}, \bibinfo
  {author} {\bibfnamefont {J.}~\bibnamefont {Gunasekera}}, \bibinfo {author}
  {\bibfnamefont {K.}~\bibnamefont {Tarwater}}, \ and\ \bibinfo {author}
  {\bibfnamefont {T.}~\bibnamefont {Heitmann}},\ }\href {\doibase
  10.1063/1.3671795} {\bibfield  {journal} {\bibinfo  {journal} {Journal of
  Applied Physics}\ }\textbf {\bibinfo {volume} {111}},\ \bibinfo {eid}
  {07D905} (\bibinfo {year} {2012})}\BibitemShut {NoStop}%
\bibitem [{\citenamefont {Katsufuji}\ \emph {et~al.}(2001)\citenamefont
  {Katsufuji}, \citenamefont {Mori}, \citenamefont {Masaki}, \citenamefont
  {Moritomo}, \citenamefont {Yamamoto},\ and\ \citenamefont
  {Takagi}}]{katsufuji01}%
  \BibitemOpen
  \bibfield  {author} {\bibinfo {author} {\bibfnamefont {T.}~\bibnamefont
  {Katsufuji}}, \bibinfo {author} {\bibfnamefont {S.}~\bibnamefont {Mori}},
  \bibinfo {author} {\bibfnamefont {M.}~\bibnamefont {Masaki}}, \bibinfo
  {author} {\bibfnamefont {Y.}~\bibnamefont {Moritomo}}, \bibinfo {author}
  {\bibfnamefont {N.}~\bibnamefont {Yamamoto}}, \ and\ \bibinfo {author}
  {\bibfnamefont {H.}~\bibnamefont {Takagi}},\ }\href {\doibase
  10.1103/PhysRevB.64.104419} {\bibfield  {journal} {\bibinfo  {journal} {Phys.
  Rev. B}\ }\textbf {\bibinfo {volume} {64}},\ \bibinfo {pages} {104419}
  (\bibinfo {year} {2001})}\BibitemShut {NoStop}%
\bibitem [{\citenamefont {Muñoz}\ \emph {et~al.}(2001)\citenamefont {Muñoz},
  \citenamefont {Alonso}, \citenamefont {Martínez-Lope}, \citenamefont
  {Casáis}, \citenamefont {Martínez},\ and\ \citenamefont
  {Fernández-Díaz}}]{munoz01}%
  \BibitemOpen
  \bibfield  {author} {\bibinfo {author} {\bibfnamefont {A.}~\bibnamefont
  {Muñoz}}, \bibinfo {author} {\bibfnamefont {J.~A.}\ \bibnamefont {Alonso}},
  \bibinfo {author} {\bibfnamefont {M.~J.}\ \bibnamefont {Martínez-Lope}},
  \bibinfo {author} {\bibfnamefont {M.~T.}\ \bibnamefont {Casáis}}, \bibinfo
  {author} {\bibfnamefont {J.~L.}\ \bibnamefont {Martínez}}, \ and\ \bibinfo
  {author} {\bibfnamefont {M.~T.}\ \bibnamefont {Fernández-Díaz}},\ }\href
  {\doibase 10.1021/cm0012264} {\bibfield  {journal} {\bibinfo  {journal}
  {Chemistry of Materials}\ }\textbf {\bibinfo {volume} {13}},\ \bibinfo
  {pages} {1497} (\bibinfo {year} {2001})},\ \Eprint
  {http://arxiv.org/abs/http://pubs.acs.org/doi/pdf/10.1021/cm0012264}
  {http://pubs.acs.org/doi/pdf/10.1021/cm0012264} \BibitemShut {NoStop}%
\bibitem [{\citenamefont {Sato}\ \emph {et~al.}(2003)\citenamefont {Sato},
  \citenamefont {Lee}, \citenamefont {Katsufuji}, \citenamefont {Masaki},
  \citenamefont {Park}, \citenamefont {Copley},\ and\ \citenamefont
  {Takagi}}]{sato03}%
  \BibitemOpen
  \bibfield  {author} {\bibinfo {author} {\bibfnamefont {T.~J.}\ \bibnamefont
  {Sato}}, \bibinfo {author} {\bibfnamefont {S.~H.}\ \bibnamefont {Lee}},
  \bibinfo {author} {\bibfnamefont {T.}~\bibnamefont {Katsufuji}}, \bibinfo
  {author} {\bibfnamefont {M.}~\bibnamefont {Masaki}}, \bibinfo {author}
  {\bibfnamefont {S.}~\bibnamefont {Park}}, \bibinfo {author} {\bibfnamefont
  {J.~R.~D.}\ \bibnamefont {Copley}}, \ and\ \bibinfo {author} {\bibfnamefont
  {H.}~\bibnamefont {Takagi}},\ }\href {\doibase 10.1103/PhysRevB.68.014432}
  {\bibfield  {journal} {\bibinfo  {journal} {Phys. Rev. B}\ }\textbf {\bibinfo
  {volume} {68}},\ \bibinfo {pages} {014432} (\bibinfo {year}
  {2003})}\BibitemShut {NoStop}%
\bibitem [{\citenamefont {Park}\ \emph {et~al.}(2003)\citenamefont {Park},
  \citenamefont {Park}, \citenamefont {Jeon}, \citenamefont {Choi},
  \citenamefont {Lee}, \citenamefont {Jo}, \citenamefont {Bewley},
  \citenamefont {McEwen},\ and\ \citenamefont {Perring}}]{park03}%
  \BibitemOpen
  \bibfield  {author} {\bibinfo {author} {\bibfnamefont {J.}~\bibnamefont
  {Park}}, \bibinfo {author} {\bibfnamefont {J.-G.}\ \bibnamefont {Park}},
  \bibinfo {author} {\bibfnamefont {G.~S.}\ \bibnamefont {Jeon}}, \bibinfo
  {author} {\bibfnamefont {H.-Y.}\ \bibnamefont {Choi}}, \bibinfo {author}
  {\bibfnamefont {C.}~\bibnamefont {Lee}}, \bibinfo {author} {\bibfnamefont
  {W.}~\bibnamefont {Jo}}, \bibinfo {author} {\bibfnamefont {R.}~\bibnamefont
  {Bewley}}, \bibinfo {author} {\bibfnamefont {K.~A.}\ \bibnamefont {McEwen}},
  \ and\ \bibinfo {author} {\bibfnamefont {T.~G.}\ \bibnamefont {Perring}},\
  }\href {\doibase 10.1103/PhysRevB.68.104426} {\bibfield  {journal} {\bibinfo
  {journal} {Phys. Rev. B}\ }\textbf {\bibinfo {volume} {68}},\ \bibinfo
  {pages} {104426} (\bibinfo {year} {2003})}\BibitemShut {NoStop}%
\bibitem [{\citenamefont {Roessli}\ \emph {et~al.}(2005)\citenamefont
  {Roessli}, \citenamefont {Gvasaliya}, \citenamefont {Pomjakushina},\ and\
  \citenamefont {Conder}}]{Roessli05}%
  \BibitemOpen
  \bibfield  {author} {\bibinfo {author} {\bibfnamefont {B.}~\bibnamefont
  {Roessli}}, \bibinfo {author} {\bibfnamefont {S.}~\bibnamefont {Gvasaliya}},
  \bibinfo {author} {\bibfnamefont {E.}~\bibnamefont {Pomjakushina}}, \ and\
  \bibinfo {author} {\bibfnamefont {K.}~\bibnamefont {Conder}},\ }\href
  {http://dx.doi.org/10.1134/1.1931017} {\bibfield  {journal} {\bibinfo
  {journal} {JETP Letters}\ }\textbf {\bibinfo {volume} {81}},\ \bibinfo
  {pages} {287} (\bibinfo {year} {2005})},\ \bibinfo {note}
  {10.1134/1.1931017}\BibitemShut {NoStop}%
\bibitem [{\citenamefont {Sekhar}\ \emph {et~al.}(2005)\citenamefont {Sekhar},
  \citenamefont {Lee}, \citenamefont {Choi}, \citenamefont {Lee},\ and\
  \citenamefont {Park}}]{sekhar05}%
  \BibitemOpen
  \bibfield  {author} {\bibinfo {author} {\bibfnamefont {M.~C.}\ \bibnamefont
  {Sekhar}}, \bibinfo {author} {\bibfnamefont {S.}~\bibnamefont {Lee}},
  \bibinfo {author} {\bibfnamefont {G.}~\bibnamefont {Choi}}, \bibinfo {author}
  {\bibfnamefont {C.}~\bibnamefont {Lee}}, \ and\ \bibinfo {author}
  {\bibfnamefont {J.-G.}\ \bibnamefont {Park}},\ }\href {\doibase
  10.1103/PhysRevB.72.014402} {\bibfield  {journal} {\bibinfo  {journal} {Phys.
  Rev. B}\ }\textbf {\bibinfo {volume} {72}},\ \bibinfo {pages} {014402}
  (\bibinfo {year} {2005})}\BibitemShut {NoStop}%
\bibitem [{\citenamefont {Lewtas}\ \emph {et~al.}(2010)\citenamefont {Lewtas},
  \citenamefont {Boothroyd}, \citenamefont {Rotter}, \citenamefont
  {Prabhakaran}, \citenamefont {M\"uller}, \citenamefont {Le}, \citenamefont
  {Roessli}, \citenamefont {Gavilano},\ and\ \citenamefont
  {Bourges}}]{Lewtas10}%
  \BibitemOpen
  \bibfield  {author} {\bibinfo {author} {\bibfnamefont {H.~J.}\ \bibnamefont
  {Lewtas}}, \bibinfo {author} {\bibfnamefont {A.~T.}\ \bibnamefont
  {Boothroyd}}, \bibinfo {author} {\bibfnamefont {M.}~\bibnamefont {Rotter}},
  \bibinfo {author} {\bibfnamefont {D.}~\bibnamefont {Prabhakaran}}, \bibinfo
  {author} {\bibfnamefont {H.}~\bibnamefont {M\"uller}}, \bibinfo {author}
  {\bibfnamefont {M.~D.}\ \bibnamefont {Le}}, \bibinfo {author} {\bibfnamefont
  {B.}~\bibnamefont {Roessli}}, \bibinfo {author} {\bibfnamefont
  {J.}~\bibnamefont {Gavilano}}, \ and\ \bibinfo {author} {\bibfnamefont
  {P.}~\bibnamefont {Bourges}},\ }\href {\doibase 10.1103/PhysRevB.82.184420}
  {\bibfield  {journal} {\bibinfo  {journal} {Phys. Rev. B}\ }\textbf {\bibinfo
  {volume} {82}},\ \bibinfo {pages} {184420} (\bibinfo {year}
  {2010})}\BibitemShut {NoStop}%
\bibitem [{\citenamefont {Vajk}\ \emph {et~al.}(2006)\citenamefont {Vajk},
  \citenamefont {Kenzelmann}, \citenamefont {Lynn}, \citenamefont {Kim},\ and\
  \citenamefont {Cheong}}]{vajk06}%
  \BibitemOpen
  \bibfield  {author} {\bibinfo {author} {\bibfnamefont {O.~P.}\ \bibnamefont
  {Vajk}}, \bibinfo {author} {\bibfnamefont {M.}~\bibnamefont {Kenzelmann}},
  \bibinfo {author} {\bibfnamefont {J.~W.}\ \bibnamefont {Lynn}}, \bibinfo
  {author} {\bibfnamefont {S.~B.}\ \bibnamefont {Kim}}, \ and\ \bibinfo
  {author} {\bibfnamefont {S.-W.}\ \bibnamefont {Cheong}},\ }\href {\doibase
  10.1063/1.2162090} {\bibfield  {journal} {\bibinfo  {journal} {Journal of
  Applied Physics}\ }\textbf {\bibinfo {volume} {99}},\ \bibinfo {eid} {08E301}
  (\bibinfo {year} {2006})}\BibitemShut {NoStop}%
\end{thebibliography}

%

\end{document}